\documentstyle[aps]{revtex}

\def \b {\begin{equation}}
\def \e {\end{equation}}

\def \l {\label}
\def \D {\mbox{D}}

\def \d {\mbox{d}}
\def \bm {\bibitem}

\begin{document}
\title{Particle number density fluctuations and pressure effects on
structure formation}
\author{D. R. Matravers$^1$\thanks{email address: david.matravers@port.ac.uk}
 and J. Triginer$^2$\thanks{email address:
 jtriginer@gcelsa.com}}
\address{~}
\address{$^{1}$Relativity and Cosmology Group, School of Computer Science
and Mathematics, University of Portsmouth,  Portsmouth~PO1~2EG,
UK}
\address{$^{2}$Departament de Fisica, Universitat Autonoma de Barcelona,
08193~Bellaterra, Spain.}
 \maketitle
\begin{abstract}
We provide a covariant and gauge-invariant approach to the
question of how a first order pressure can be incorporated
self-consistently in a cosmological scenario.  The approximation
is relevant, in the linear regime, to weakly self-interacting or
warm dark matter models. We also derive number density
fluctuations in which new modes appear because the number density
fluctuations are no longer simply proportional to the density
fluctuations.\\
\end{abstract}


\section{Introduction}

According to current cosmological models, the large-scale
structure, as seen today, formed from the evolution of
gravitational instabilities which originated from inflationary
processes in the primordial fluid. In studying the evolution of
such instabilities, analytically in the fluid
approximation\footnote{For high accuracy the full Boltzmann
equations are used and these require numerical codes, which
provide different insights into the process.}, one usually assumes
a dust or radiation energy momentum tensor. The object of this
paper is to consider some implications of modifying these
assumptions.

First we note that while the two extreme fluid states are
reasonable for most of the history of the universe, there are
short, but at the same time important, specific periods in which
matter is in the non-relativistic regime with a tiny although
non-negligible pressure.  These are periods during which
collisions, or small random motions (velocity dispersion) of
collisionless matter, give rise to a small but non-zero matter
pressure. Even though in these cases the matter pressure in the
background may be negligible, small disturbances (perturbations)
of the spacetime induce a tiny pressure term which can modify the
evolution of the inhomogeneities.  Such a scenario may also have
applications  when studying weakly self-interacting matter (for
details see Spergel and Steinhardt \cite{fh}) or warm dark matter
(for more details of this approximation see Hogan and Dalcanton
\cite{fh}) in the linear regime.  These dark matter models have
been the subject of much recent work using numerical integrations
in the nonlinear regime, where they have been employed to
investigate ways of resolving problems with the standard CDM model
on small scales (see \cite{fh}) in galaxy formation.

Second we take the particulate nature of matter seriously.  It
must be born in mind that in structure formation one really needs
to look at the clumping of a set of particles rather than at the
evolution of energy density instabilities. Although in taking the
dust assumption to be valid such a distinction is not relevant,
this is not true for those periods in which matter is undergoing a
transition to the non-relativistic regime and, consequently, for
which the dust approximation is not accurate enough, even in the
linear regime. To model the effects of the particulate nature of
the matter we make use of thermodynamic arguments. An alternative
Kinetic Theory approach is given in \cite{MT2}.

The method we use is based on the Ellis-Bruni \cite{EB} covariant
and gauge-invariant perturbation formalism, which has proved
useful in different cosmological problems  \cite{MT2}, \cite{MT},
\cite{varis}. It is one of the gauge-independent descriptions of
relativistic perturbations, described in detail in the literature,
see for instance
\cite{EB},\cite{Stewart},\cite{Bardeen},\cite{maj}.

In our treatment of the fluid models we generalise the usual
equations for the growth of cosmological perturbations for matter
in a Friedman-Robertson-Walker (FRW) background to include first
order pressure effects due to a non-relativistic component, in the
radiation and in the matter dominated eras. Thus the simplifying
hypothesis that the matter is dust, i.e. has vanishing pressure,
is relaxed, in both cases, to include a small pressure due to
collisions or thermal motions of the particles which make up the
fluid.

\section{Matter versus energy density perturbations}
The starting point for the the Ellis-Bruni \cite{EB} formalism is
the choice of a time-like congruence $u^a$ along which typical
(comoving) observers move. Once this four-velocity has been
selected all the physical quantities are split into their spatial
and time-like parts using the spatial projector tensor $h^{ab}$.
The relevant physical quantities are defined as follows
\b
\delta_a=a\frac{\D_a\rho}{\rho},\ \ \theta_a=a\D_a\theta,\ \
p_a=a\frac{\D_a p}{\rho},\ \ \nu_a=a\frac{\D_an}{n}, \l{1}
\e
where $\D_a\equiv h^b_a\nabla_b$ is the spatial projection of the
covariant derivative, $\theta\equiv u^a_{;a}$ the expansion, $p$
the pressure, $\rho$ the energy density and $a$ is a
representative or average length scale defined by $3\dot{a}/a =
\theta$. In a FRW metric $a$ is the background scale factor and we
will use the notation $3 H = \theta$ in the background. The
quantities $\delta_a$, $p_a$, $\theta_a$ and $\nu_a$ describe the
spatial inhomogeneity of quantities which are homogeneous and
isotropic in the background. Each of them is spatial (contraction
with $u^a$ vanishes), covariant and gauge-invariant to first
order, as all of them vanish in the FRW background \cite{Stewart}.
The scalar parts of perturbations described in (\ref{1}) are
obtained by taking the comoving spatial divergence, e.g.,
$\delta=a\D^a\delta_a$.

The general procedure to obtain evolution and constraint equations
for the quantities defined in (\ref{1}) starts from a suitable
spatial projection of the Ricci and Bianchi identities as given in
terms of the fluid variables once the field equations have been
taken into account (see for example \cite{cargese} for a
pedagogical account). Here we shall follow the notation of
\cite{MT}.

Restricting ourselves to a spatially flat FRW background, the
equations governing the evolution of the comoving fractional
density gradient $\delta_a$ and the comoving expansion gradient
$\theta_a$ are (see \cite{MT} equations 25,26)

\begin{eqnarray}
&&\dot{\delta}_a=3wH\delta_a-(1+w)\theta_a,\l{2}\\
&&\dot{\theta}_a-\frac{9}{2}H^2p_a+2H\theta_a-a\D_a\D^b\dot{u}_b
+\frac{3H^2}{2}(\rho_a+3p_a)=0,\l{3}
\end{eqnarray}
where
\b
w=\frac{p}{\rho}\,. \l{4}
\e
Note that equation (\ref{2}) omits second order terms proportional
to $\sigma_{ab} \delta^{b}$ and $\omega_{ab} \delta^{b}$ where
$\sigma_{ab}$ and $\omega_{ab}$ are the shear and vorticity
tensors.

We will now discuss perturbations in the particle number density.
This discussion is motivated by the idea that a full understanding
of large scale structure formation in the framework of
gravitational instabilities requires an analysis of how particles,
which comprise the cosmic fluid, clump together due to their
mutual attraction. Although dealing with the energy density is the
usual procedure, we believe that it is desirable to consider the
particle number density $n$ because, ultimately, particles are the
building blocks of all the cosmic structures we see today. For
dust, it does not make any difference whether one chooses $\rho$
or $n$ because no random motions are present and the internal
energy vanishes identically, yielding $\rho=mn$, i.e.
\b
\delta_a = \nu_a. \l{5}
\e

The situation is quite different when internal energy associated
with thermal motions is introduced.  The behaviour of $\nu_a$
requires a separate investigation. In \cite{MT} it is shown that
from the Gibbs equation
\b
T\d s=\d\left(\frac{\rho}{n}\right)+p\d\left(\frac{1}{n}\right),
\l{6}
\e
it follows that

\b
e_a=\delta_a-(1+w)\nu_a, \l{7}
\e
where
\b
e_a=\frac{anT\D_as}{\rho}, \l{8}
\e
is a dimensionless entropy gradient. Perturbations which are
non-dissipative ($\dot{s}=0$) and without spatial variation of the
entropy ($e_a=0$) are called isentropic in \cite{MT} because they
have the same entropy at all the points of the spacetime. In the
weaker case where no spatial variation of the entropy is allowed
$(e_{a} = 0)$ the number density perturbations are related
algebraically to the energy density perturbations by
\b
\nu_a=\frac{\delta_a}{1+w}. \l{9}
\e
In fact, $e_a=0$ is a strong assumption which does not hold in
general. Its departure from zero affects the growth of
perturbations. From the particle conservation equation
\b
\dot{n}+n\theta=0, \l{10}
\e
we get an evolution equation for $\nu_a$
\b
\dot{\nu}_a+\theta_a-\frac{\theta}{1+w}p_a=0. \l{11}
\e
If we use the standard, but generally unphysical, equation of
state $w = {\rm constant}$ then together with (\ref{2}), the
integral of (\ref{11}) yields
\b
\nu_a=\frac{\delta_a}{1+w}+k_a,\ \ \ \dot{k}_a=0. \l{13}
\e
Thus, in general, when this simple equation of state is used, a
new stationary (entropy) mode appears in the evolution of $\nu_a$
which could have an impact on the structure formation at certain
stages of the evolution of inhomogeneities. The problem with
(\ref{13}) is that it makes sense only for the very special, but
widely used, cases of dust and radiation.  In more general
circumstances such as in self interacting dark matter or where
there are collisions \cite{fh}, the equations of state should
involve two independent thermodynamical quantities, i.e. $n$ and
$\rho$. To deal with the inhomogeneities that arise in
non-relativistic matter with a non-vanishing pressure we may
follow two different approaches.\\

(i) If we assume that the fluid has reached a collision-dominated
equilibrium\footnote{For an example of such a model see the paper
by Spergel and Steinhardt  \cite{fh} where it is pointed out that
under certain circumstances the cold dark matter behaves as a
collision dominated gas} , we learn from kinetic theory
 that the equation of state in the non-relativistic regime is \cite{wei}
\b
\rho=mn+\frac{3}{2}p, \l{14}
\e
which together with the energy balance equation
\b
\dot{\rho}+ \theta(\rho+p)=0, \l{15}
\e
and the particle number balance equation (\ref{10}) leads to
\b
\dot{p}=- \frac{5}{3} \theta p, \l{16}
\e
with solution
\b
p=p_0\left(\frac{a_0}{a}\right)^5. \l{17}
\e
 Also from (\ref{14}) we get the equation
\begin{equation}
\delta_a  = \frac{mn}{\rho}\nu_a+\frac{3}{2}p_{a} \l{18}
\end{equation}
which, together with (\ref{11}) and the conservation equations,
determines $\nu_a$.\\

(ii) Without imposing the restriction of a collision-dominated
equilibrium fluid, we could tackle the problem of a collisionless
gas in which a small pressure arises because of velocity
dispersion (random motions) of the particles making up the
system\footnote{Such a dark matter model has recently been
described by Hogan and Dalcanton \cite{fh}}. In \cite{MT2} it was
shown that by neglecting higher powers of the velocity dispersion
in the Boltzmann equation, the evolution equation for pressure is
given by (\ref{16}). We remark that in obtaining this equation no
assumption about the equation of state was made.  Only the absence
of collisions and the smallness of the velocity dispersion were
used. Despite the evident mathematical similarities to case (i), a
crucial difference emerges here: for a collisionless gas the Gibbs
equation (\ref{6}) does not apply and, as we shall show below,
this leads to some new results.

In the next section we discuss cosmological settings where the
small pressure terms introduced above are included.

\section{Non-relativistic matter perturbations with pressure}

In this section we investigate the role played by a first order
pressure in the non-relativistic regime on an FRW background
filled, in the first case, by a dust gas and, in the second, by
dust and radiation. The point is that while the idealized
background can be described by dust or dust and radiation, as soon
as the dust is disturbed (by velocity dispersion or collisions) a
small nonzero pressure arises. This means that the pressure is a
first order quantity, as it vanishes in the background. This
differs from the usual approach which takes an identically zero
pressure in the background and in the real (perturbed) spacetime.
We relax this assumption by allowing first order corrections to
pressure due either to the fluid being in a collision-dominated
regime (case (i) above) or through velocity dispersion without
collisions (case (ii) above). This enables us to initiate a
theoretical discussion of the effects being studied mainly
qualitatively or numerically in the papers \cite{fh}.
One-component and two-component models will be considered in turn.
We note here that the perturbations will involve perturbations of
the entropy.  The precise relation of the covariant description of
entropy perturbations to the metric description \cite{pee} is not
straightforward and will not be discussed here.
\subsection{Dust background}

For simplicity we use an Einstein-de Sitter background. As stated
above we focus on perturbations for which pressure is a first
order quantity. From the previous section we have that, both in
case (i) and case (ii), $p$ is given by the evolution equation
(\ref{16}) leading to (\ref{17})
\b
p=p_0\left(\frac{a_0}{a}\right)^5.
\e
As $p$ is now a first order quantity we get from (\ref{2}) and
(\ref{3}),
\b
\ddot{\delta}_a-2H\dot{\delta}_a+\frac{3}{2}H^2\delta_a+\frac{a}{\rho}
\D_a\D^2p=0, \l{19}
\e
where the momentum balance equation
\b
(\rho+p)\dot{u}_a=-\D_ap, \l{20}
\e
has been used. Note that for large-scale perturbations the
Laplacian term is negligible and we recover the standard evolution
equation for dust. Hence, as expected, pressure plays a role only
at small scales.
 \subsection{Collision dominated fluid}

At this point we distinguish between the two cases discussed in
section two. If we assume that the fluid is in collision-dominated
equilibrium, then the Gibbs equation holds which, together with
the equation of state (\ref{14}), leads to
\b
\D_ap=0, \l{21}
\e
at first order, reducing (\ref{19}) to the standard equation valid
for dust at any scale. Expression (\ref{21}) implies that the
spatial gradient of $p$ is a gauge-invariant second order quantity
since it vanishes at first order. This allows us to go beyond
first order and derive a second order equation for the pressure as
we will now show. A fully covariant and gauge-invariant treatment
for second order perturbations is not possible at the moment. The
point is that in order to construct $n$-order gauge-invariant
variables following Ellis-Bruni, we have to ensure that they are
constant (usually zero) at all orders below $n$ \cite{Bruni1}. In
particular it is not clear how to find (if at all possible) a
second-order meaningful variable for the perturbed density in such
a way that it vanishes at zero and first orders. Bruni and
co-workers provide a systematic method to tackle relativistic
perturbations beyond the linear order, but one has to pay the
price of losing the gauge-invariant character \cite{Bruni2}.

It is possible, however to obtain a second order gauge invariant
constraint on the perturbations from the above results. Operate on
equation (\ref{16}) with $a \D_{a}/\rho$ and use the identity
(\ref{25})\footnote{Since $p_{a}$ is second order, the first order
identity gives rise to second order terms and any corrections are
of higher order.} below to obtain
\b
\dot{p}_{a} + p_{a} \frac{\dot{\rho}}{\rho} +
\frac{5}{3}\theta_{a} \frac{p}{\rho} + \frac{5}{3} \theta p_{a} =
0 \label{pa}
\e
and then use,
\begin{eqnarray*}
\dot{\delta}_{a} & = & - \theta_{a} + O(2) \\
 \dot{\rho}& = & - 3 \left(\frac{\dot{a}}{a}\right)p + O(1) \\
 \end{eqnarray*}
 derived from equations ({\ref{2}) and (\ref{15}) and where $O(n)$
 indicates quantities of order $n$.  Note that $p_{a}$ is of order
 $O(2)$.
 After a change in the independent
 variable the equations (\ref{pa}) become
\b
p_a^{\prime}+
2\left(\frac{a_{0}}{a}\right)p_a=\frac{5p}{3\rho}\delta_a^{\prime}\,
, \l{22}
\e
where a prime denotes a derivative with respect to $a/a_{0}$.
According to the argument above and the fact that $p$ and
$\delta_a$ are first order quantities, equation (\ref{22})
contains only second order quantities. It is one of the set of
second order equations. We treat it as a constraint on the gauge
invariant pressure perturbations in the sense that a full solution
to all the second order perturbation equations will have to
satisfy (\ref{22}) although the actual solution may be more
restricted. With this in mind we proceed to solve it on a dust
background. In equation (\ref{22}) we use equations (\ref{17}) and
(\ref{28}), below, for the dust background and the standard first
order solution for dust
\b
\delta_a = K_{a}^{+}\left(\frac{a}{a_{0}}\right) +
K_{a}^{-}\left(\frac{a}{a_{0}}\right)^{-3/2}, \l{23}
\e
where $K_{a}^{+}$ and $K_{a}^{-}$ are constants of order $O(1)$,
to obtain
\b
p_a= A_{1} K_{a}^{+}\left(\frac{a_{0}}{a}\right) + A_{2}
K_{a}^{-}\left(\frac{a_{0}}{a}\right)^{7/2} +
K_{a}^{*}\left(\frac{a_{0}}{a}\right)^{2}, \ \ \ \ A_{1} \equiv
\frac{5p_{0}}{ 3\rho_{0} a_{0}} \equiv A_{2} a_{0}^{5/2}. \l{24}
\e
Here $A_{1}$, $A_{2}$, $K_{a}^{+}$ and $K_{a}^{-}$ are constants
of order $O(1)$, so the products are of order $O(2)$ and
$K_{a}^{*}$ are constants of order $O(2)$. The solution (\ref{24})
is a simple and exact solution giving a limitation on the
evolution of a second order quantity in the Ellis-Bruni formalism.
The point is that we can apply the formalism because $\D_ap$
vanishes at the zero and at the first order approximation.
\subsection{Collision free fluid}

The more interesting and physically appealing situation is that of
a collisionless gas for which the foregoing arguments are not
applicable. Applying the operator $a\D_a\D^2$ to (\ref{16}) and
using the first order identity
\b
(a\D_af)^{\displaystyle{\cdot}}=a\D_a\dot{f}, \l{25}
\e
where $f$ denotes any first order scalar or tensor quantity, we
get
\b
(a\D_a\D^2p)^{'}=- 7 \left(\frac{a_{0}}{a}\right)a\D_a\D^2p,
\l{26}
\e
and hence
\b
a\D_a\D^2p=A_a^0\left(\frac{a_0}{a}\right)^7,\ \ \ \ A_a^0\equiv
a\D_a\D^2p|_{a=a_0},\ \ \ \dot{A}_a^0=0. \l{27}
\e

For a dust background
\b
H=\sqrt{\frac{\rho_0}{3}}\left(\frac{a_0}{a}\right)^{3/2},\ \ \
\rho=\rho_0\left(\frac{a_0}{a}\right)^{3}, \l{28}
\e
so equation (\ref{19}) reduces to
\b
\delta_a^{''} + \left(\frac{H^{'}}{H} +
3\frac{a_{0}}{a}\right)\delta_a^{'} - \frac{3}{2}
\left(\frac{a_{0}}{a}\right)^{2}\delta_a =
\frac{A_a^0}{\rho_0H^2}\left(\frac{a_0}{a}\right)^6, \l{29}
\e
which can be solved to give
\b
\delta_a=K_{a}^{+}\left(\frac{a}{a_0}\right) +
K_{a}^{-}\left(\frac{a_0}{a}\right)^{3/2} -
\frac{3A_a^0}{\rho_0^2}\left(\frac{a_0}{a}\right) \l{30}
\e
where $\dot{K}^{\pm}_{a} = 0$.  Thus a new non-adiabatic mode is
obtained which decays, albeit at a slower rate than the standard
decaying mode for dust.  Although the effect of the mode will be
at small scales, as desired, the fact that it is decaying means
that it is unlikely to be significant.

\subsection{Dust-radiation background}

The novelty in this subsection is that the dynamics of the
background is given by a decoupled mixture of dust and radiation.
We will consider perturbations of the matter component which again
will be taken as dust in the background but with a non-vanishing
first order pressure in the real spacetime.   This situation
mimics that of the cosmological fluid around the time of the
transition from the radiation to the matter dominated era with a
matter component which is acquiring a major role in the dynamics.
In the following $p_r$, $\rho_r$ and $\rho_m$ denote the zero
order pressure and energy density of radiation and matter
respectively, whereas $p_m$ is the first order pressure of matter
as given by (\ref{17}).

Assuming that the fluids are decoupled and share the same
4-velocity\footnote{Using the same velocity is valid when the two
fluids are coupled.  Here it is a simplifying assumption which
enables us to provide an analytical treatment of a problem, that
has only been treated numerically up to now, and we are able to
obtain new results. The role of the assumption is made explicit,
in particular in the comment following equation (\ref{39}), where
it is relevant.} we get
\begin{eqnarray}
&&(\rho_r+p_r)\dot{u}_a+\D_ap_r=0,\l{31}\\
&&\rho_m\dot{u}_a+\D_ap_m=0.\l{32}
\end{eqnarray}
From these equations and the equation of state for radiation we
find that
\b
\frac{\D_ap_r}{\rho_r+p_r}=\frac{\D_ap_m}{\rho_m}\Rightarrow
\delta_a^r=4p_a^m,\l{33}
\e
where
\b
\delta_a^r\equiv a\frac{\D_a\rho_r}{\rho_r},\ \ \ p_a^m\equiv
a\frac{\D_ap_m}{\rho_m}. \l{34}
\e
Note that if we were to keep the matter component as dust both in
the background and in the real spacetime, we would have ended up
with $\dot{u}_a=0=\D_ap_r$, i.e. radiation perturbations would not
have been allowed.\\

In order to get an evolution equation for $\theta_a$ we start with
the Raychaudhuri equation
\b
\dot{\theta}+\frac{1}{3}\theta^2-\D^a\dot{u}_a+\frac{1}{2}(\rho+3p)=0,
\l{35}
\e
where $\rho=\rho_r+\rho_m$ is the total energy density and
$p=p_r+p_m$ is the total pressure. We have to be very careful in
dealing with $p$ since in the background $p=p_r$ but when we apply
the spatial derivative operator we get $\D^ap=\D^ap_r+\D^ap_m$.
Bearing all this in mind the evolution equation for $\theta_a$
becomes
\b
\dot{\theta}_a + 2H\theta_a =  a\D_a\D^b\dot{u}_b -
\frac{9}{2}H^2a\dot{u}_a(1+w) - \frac{1}{2}
(8\rho_rp_a^m+3\rho_mp_a^m+\rho_m\delta_a^m), \l{36}
\e
where equation (\ref{33}) has been used.

Before dealing with the acceleration term we recall the identity
\cite{MT}
\b
\D^2(\D_{a}f)=\D_a(\D^2f)+\frac{2}{3}(\rho-3H^2)\D_af+2\dot{f}\mbox{curl}\
\omega_a. \l{37}
\e
From this identity and bearing in mind that $p_m$ is a first order
quantity, we have
\b
a\D_a\D^b\dot{u}_b=-a\frac{\D_a\D^2p_m}{\rho_m}=-a\frac{\D^2\D_ap_m}{\rho_m}=
-\D^2p_a^m. \l{38}
\e
It is important to note that the result obtained from commuting
the operators $\D_a$ and $\D^2$ would have been different if we
had used $p_r$ instead of $p_m$. In the latter case a first order
term $\sim\dot{\rho}_r\mbox{curl}\ \omega_a$ would have arisen.
This means that consistency requires
\b
\mbox{curl}\ \omega_a=0, \l{39}
\e
i.e., {\em in a radiation-matter decoupled mixture for which (a)
both fluids share the same 4-velocity and (b) the matter pressure
arises from perturbations so that $p_m$ is a first order quantity,
${\rm curl}\ \omega_a$ vanishes at first order}.\\

Changing the independent variables from $t$ to $(a/a_{0})$, we get
the evolution equation for $\theta_{a}$,
\b
\theta'_a + 2\theta_a \left(\frac{a_{0}}{a}\right) =
\left(\frac{a_{0}}{a}\right)\left\{\frac{9H}{2}\left(4+\frac{3\rho_m}{\rho_r}
\right)p_a^m - \frac{\D^2p_a^m}{H} - \frac{1}{2H}
\left[p_a^m(8\rho_r+3\rho_m) + \rho_m\delta_a^m\right]\right\}.
\l{40}
\e
The background equations for dust plus radiation are
\b
\rho=\rho_m+\rho_r=\frac{3}{\beta a^4}(1+a\alpha),\ \ \ p=p_r=
\frac{1}{\beta a^4}, \ \ \ H^2=\frac{1}{\beta a^4}(1+a\alpha),
\l{41}
\e
where
\b
\beta \equiv \frac{3}{\rho_0^ra_0^4},\ \ \ \
 \alpha \equiv \frac{\rho_0^m}{\rho_0^ra_0}. \l{42}
\e
From the evolution equation for $p$ for a collisionless gas
\b
\dot{p}_m=-\frac{5}{3}\theta p_m, \l{43}
\e
and using the equation (\ref{25}) we have
\b
\dot{p}_a^m=-2Hp_a^m\Rightarrow (p^m_a)'= - \frac{2}{a}p_a^m,
\l{44}
\e
which gives
\b
p_{a}^{m} = K_{a}^{m}\left(\frac{a_0}{a}\right)^{2},\ \ \
\dot{K}^{m}_a=0. \l{45}
\e
Using similar reasoning for $\D^2p_a^m$ and with the help of the
identity
\b
\left(\D^2 f\right)^{.}=\D^2\dot{f}-2H\D^2f+\dot{f}\D^a\dot{u}_a,
\l{46}
\e
from \cite{MT}, we get
\b
\D^2p_a^m=M_{a}^{m} \left(\frac{a_0}{a}\right)^4,\ \ \ \
\dot{M}_{a}^{m}=0. \l{47}
\e
 We now define the scalar parts of the perturbations by
 \b
 \delta \equiv a\D^{a} \delta_{a}, \ \ Z
 \equiv a\D^{a} \theta_{a}. \label{sc}
 \e
 From (\ref{45}) it follows that
 \b
 a\D^{a} p_{a}^{m} = K\left(\frac{a_{0}}{a}
 \right)^{2}, \ \ \dot{K} = 0,
 \e
 where $K = a\D^{a}K_{a}^{m}$, and that
 \b
 a\D^{a}[a\D_{a}(\D^{b} \dot{u}_{b})] = - M\left(
 \frac{a_{0}}{a} \right)^{4}, \ \ \dot{M} = 0,
 \e
 where $M \equiv a\D^{a} M_{a}^{m}$.  From equations (\ref{32})
 and (\ref{33}) it follows that
 \b
 a\D^{a}(a \dot{u}_{a}) = - K \left(\frac{a_{0}}{a} \right)^{2}.
 \e
 Operating on equation (\ref{36}) with $a\D^{a}$ gives
 \b
 \dot{Z} + 2 H Z + \frac{1}{2}\rho_{m} \delta = - M \left(
 \frac{a_{0}}{a}\right)^{4} - K \left(\frac{a_{0}}{a}
 \right)^{2} [4 \rho_{r} + \frac{3}{2}\rho_{m} - \frac{9}{2}
 (1 + w)H^{2}] . \label{B}
 \e
 For the matter scalar perturbations, $w = 0$ and equation (\ref{2})
 implies $Z = - \dot{\delta}$.  The evolution equation for $\delta$
 written in terms of the scaled independent variable
  $$
  x = \alpha a
 = \frac{\rho^{m}_{0}}{\rho^{r}_{0}}\frac{a}{a_{0}}
  $$
 is
\b
\delta^{''} + \frac{3x+2}{2x(x+ 1)} \delta^{'}
-\frac{3}{2}\frac{1}{x(x+ 1)} \delta  = F(x), \l{48}
\e
where the prime now denotes a derivative with respect to the new
independent variable $x$ and
 \b
 F(x) = \frac{3}{1 + x} \left[\frac{M}{\rho_{0}^{r} x^{2}} +
 \frac{5}{2} K
 \left(\frac{\rho_{0}^{m}}{\rho_{0}^{r}}\right)^{2}\frac{1}{x^{4}}
  \right]
 \e
is the inhomogeneous part. It is immediately apparent that
 \b
 \delta_{1} = 3x + 2 \label{sol1}
 \e
  is a solution to the homogeneous part of
equation (\ref{48}).  The method of reduction of order \cite{km}
then leads to a complete solution.  The second solution of the
homogeneous part is
\b
 \delta_{2} = \delta_{1} \int \frac{dx}{(3x + 2)^{2} x \sqrt{x +
 1}}\, ,
\e
which can be integrated using partial fractions to obtain
 \b
 \delta_{2} = \frac{3}{2}\sqrt{x + 1} + \frac{(3x + 2)}{4}
 \ln \left[\frac{\sqrt{x + 1} - 1}{\sqrt{x + 1} + 1}
 \right] \label{gp}.
 \e
 The homogeneous form of equation (\ref{48}) was obtained by
 M\'esz\'aros \cite{mes} in 1974 who obtained a solution by transforming
 the equation to a hypergeometric form. Groth and Peebles \cite{gro} obtained
 the solutions (\ref{sol1} , \ref{gp}) and gave the asymptotic
 behaviour
 \[
   \begin{array}{llllll}
   \delta_{2} & \approx &  \frac{1}{2}\ln(\tau/8), \ \ \ \ \ & x & \leq &
   1,
   \\
   \delta_{2} & \approx & - 8/(45 \tau), & x & \geq & 1,
   \end{array}
   \]
   where
   $$
   \tau  =  \frac{1}{\sqrt{3}}\frac{(\rho^{m})^{2}}{(\rho^{r})^{3/2}}
   t.
   $$
      The results are further described in Peebles's book \cite{peb}.
   \\

The full solution to the inhomogeneous equation (\ref{48}) is
given by
\b
\delta = C_{1} \delta_{1} + C_{2} \delta_{2}  + \delta_{1} \int
\frac{1}{E \delta_{1}^{2}} \left(\int E \delta_{1}F dx\right)dx,
\label{50}
\e
where $E = \exp \left(\int (3x + 2)[2x(x + 1)]^{-1}dx\right) = x
\sqrt{x + 1}$.

The expression $E \delta_{1}F$ can be written in the form
\b
E \delta_{1}F = \frac{1}{\sqrt{x + 1}}\left\{A_{0} +
\frac{A_{1}}{x} + \frac{A_{0}}{3 x^{2}} + \frac{A_{1}}{3 x^{3}}
\right\} \l{pre}
\e
where $A_{0} = 9 M/\rho^{r}_{0}$, $A_{1} = \frac{45}{2} K
\left(\rho_{0}^{m}/\rho_{0}^{r}\right)^{2}$. From this the
particular integral can be calculated term by term. After
integration, the first term $\delta_{p1}$ of the particular
integral is given by
 \begin{eqnarray*}
 \delta_{p1} & = & \delta_{1}\int\frac{1}{E \delta_{1}^{2}}\left(\int
 \frac{A_{0}}{\sqrt{x + 1}}\, dx \right) \, dx \\
 & = &  A_{0} + \frac{A_{0}}{2}(3x + 2) \ln \left(\frac{x}{3x + 2}
 \right).
 \end{eqnarray*}

\noindent The remaining terms are more difficult to integrate.
From the first term it appears that the full inhomogeneous
solution will contribute new modes. Among these it will contribute
a constant mode and may contribute a mode similar to the second
term in the solution $\delta_{2}$ and hence modify the coefficient
of the factor $(3x + 2)$ which could affect the asymptotic
behaviour. The significance of this has to be viewed with caution
because the remaining, still to be determined terms, may
contribute further modes or cause cancellations. From the physics
one is led to expect that the contribution of a small pressure
perturbation will only be decaying modes.

Two other features of equation (\ref{48}) are worthy of note.
First, for large $x$ the asymptotic form of the homogeneous
equation is the usual equation for first order perturbations in
dust.  Second, even for initial conditions in which $\delta$ and
its first derivative with respect to $x$ are zero, perturbations
will arise from the influence of the function $F(x)$. In other
words small pressure effects or number density fluctuations can
source density perturbations. This is possibly more significant,
for the formation of structure, than the modification of the
modes.
 \section{Conclusion}

For analytical as opposed to numerical modelling of the evolution
of inhomogeneities in the universe, it is conventional to use a
fluid approximation.  At a detailed level this is at variance with
reality because the matter is generally more particulate than
hydrodynamics allows.  Also most of the analytical literature
assumes that the cosmological non-relativistic matter is dust.  If
we take the particulate nature into account,  then at early times
weak self-interactions or small random motions of collisionless
matter give rise to a small but non-zero pressure in a fluid
model. With this motivation we have discussed some of the
implications of including a first order pressure in the matter
distribution for fluid models.

We have derived the relation between the energy density
perturbations and the number density perturbations for the
isentropic case and we show that in the simple non-isentropic case
with $p/\rho = {\rm constant}$ a new stationary mode appears. This
new mode will affect the formation of structure at certain stages
in the evolution of the inhomogeneity.

In the main body of the paper we derive an exact solution to the
energy density perturbation equation when the matter pressure is
non-zero at first order.  The implications of assuming a non-zero
first order matter pressure are quite deep as is illustrated in
equation (\ref{24}), which is an equation for second order gauge
invariant quantities, and in equation (\ref{32}), which together
with (\ref{31}) shows that if $\D_{a}p_{m}$ were equal to $0$,
then $\dot{u}_{a} = 0 = \D_{a}p_{r}$. This would reduce the
problem to the standard case.

The complete solution (\ref{50}) is different from the usual
solutions used in CDM approximations and from the M\'esz\'aros
solution. Having such a solution may be useful as an analytical
tool for understanding some features of recent numerical work on
weakly self-interacting or warm dark matter models in the
non-linear regime \cite{fh}. This is a subject of further
investigation.\\

\noindent {\bf Acknowledgements}\\

\noindent The authors thank Roy Maartens and Marco Bruni for
useful discussions on this work. J. Triginer was supported in part
by a Royal Society Joint Project Grant.

\end{document}